\documentclass[12pt]{article}
\usepackage{amssymb,amsmath,epsfig}

\begin{document}

\title{\bf Dynamics of Non-adiabatic Charged Cylindrical Gravitational Collapse}
\author{M. Sharif \thanks {msharif.math@pu.edu.pk} and G. Abbas
\thanks {abbasg91@yahoo.com}\\
Department of Mathematics, University of the Punjab,\\
Quaid-e-Azam Campus, Lahore-54590, Pakistan.}

\date{}

\maketitle
\begin{abstract}
This paper is devoted to study the dynamics of gravitational
collapse in the Misner and Sharp formalism. We take non-viscous heat
conducting charged anisotropic fluid as a collapsing matter with
cylindrical symmetry. The dynamical equations are derived and
coupled with the transport equation for heat flux obtained from the
M$\ddot{u}$ller-Israel-Stewart causal thermodynamic theory. We
discuss the role of anisotropy, electric charge and radial heat flux
over the dynamics of the collapse with the help of coupled equation.\\
\end{abstract}

\section{Introduction}

One of the most important problems in the gravitation theory and
relativistic astrophysics is to understand the end state of a
continual gravitational collapse. A massive star undergoes to
gravitational collapse at the end of its life cycle. This happens
when all the internal nuclear forces fail to supply the sufficiently
high pressure to counter-balance gravity. The compact objects such
as white dwarfs, neutron stars and black hole are the results of
possible stages of the collapsing astronomical objects. In white
dwarfs and neutron stars, gravity is neutralized by electron and
neutron degeneracy pressure respectively and black hole is a
complete collapsed object.

Oppenheimer and Snyder (1939) are the pioneers who studied
gravitational collapse of an adiabatically flowing dust. This was
idealized problem because dust is unrealistic matter and one cannot
ignore the effects of pressure on the formation of spacetime
singularity. A more analytic analysis was made by Misner and Sharp
(1964) with perfect fluid in the inner region of a star. They
formulated the dynamical equations governing adiabatic relativistic
collapse. In both cases, vacuum was taken in the exterior region of
a star.

The concept of non-vacuum exterior of a star was introduced by
Vaidya (1951) for the radiating fluid in the interior region of the
star. Goswami (2007) formulated a more realistic collapsing model by
taking the radiating dust matter in the interior of a star. He
remarked that bounce in the collapse is due to the dissipation.
Debnath et al. (2005) explored gravitational collapse of the
non-adiabatic fluid by assuming quasi-spherical Szekeres spacetime
in the interior and plane symmetric Vaidya solution in the exterior
region. By using the local conservation of momentum, they studied
the thermodynamical behavior of the collapsing matter.

An extensive literature survey (Herrera et al. 2004 and  Mitra
2006)predicts that gravitational collapse is highly dissipative
process. This indicates that the effects of the dissipation must be
included in the study of collapse for its better understanding.
Herrera and Santos (2004) explored dynamical description of
gravitational collapse by using Misner and Sharp's formulation.
Matter under consideration was distributed with spherically
symmetric and energy loss in the form of heat flow and radiation.
Chan (2001) studied the realistic model of radiating star which
undergoes dissipation in the form of radial heat flow and shear
viscosity. Herrera et al. (2009) also formulated the dynamical
equations by including dissipation in the form of heat flow,
radiation, shear and bulk viscosity and then coupled with causal
transport equations with spherical symmetry. Herrera (2006)
discussed the inertia of heat and its role in the dynamics of
dissipative collapse with outgoing radial heat flux by using
spherical symmetry.

Most of the work available in spherical symmetry is due its simplest
symmetry. To generalize the geometry of the star, people worked on
gravitational collapse using the non-spherical symmetry. The
existence of cylindrical and plane gravitational waves provides
strong motivation in this regard. Herrera et al. (2005) formulated
the set of equations with regularity and matching conditions for the
static cylindrically symmetric distribution of matter. Sharif and
Ahmad (2007) studied cylindrically symmetric gravitational collapse
of two perfect fluids using the high speed approximation scheme.
They investigated the emission of gravitational radiations from
cylindrically symmetric gravitational collapse. Nolan (2002)
investigated naked singularities in the cylinderical gravitational
collapse of counter rotating dust shell. Di Prisco et al. (2009)
discussed the shear free cylindrical gravitational collapse by using
junction conditions. Nakao et al. (2009) studied gravitational
collapse of a hollow cylinder composed of dust. Recently, Sharif and
Rehmat (2010) discussed the dynamics of viscous dissipative plane
symmetric gravitational collapse.

The behavior of electromagnetic field in gravitational field has
been the subject of interest for many people. Thorne (1965)
developed the concept of cylindrical energy and investigated that a
strong magnetic field along the symmetry axis may halt the
cylindrical collapse of a finite cylinder before it reached to
singularity. In recent papers (Sharif and Abbas 2009, 2010a, 2010b),
we have studied the effects of the electromagnetic field on the
gravitational collapse by taking the homogenous, non-homogeneous and
spherical model. Di Prisco et al. (2007) derived the dynamical
equations for the spherically symmetric collapse by including
electromagnetic field. This work has been extended by Sharif and
Siddiqa (2011) for the charged plane symmetric gravitational
collapse. Also, Sharif and Fatima (2011) discussed dynamics of
adiabatic charged viscous cylindrical gravitational collapse.

This paper is aimed to study the dynamics of non-adiabatic charged
cylindrically symmetric gravitational collapse to see the effects of
charge and heat flux on the process of collapse. The plan of the
paper is the following. In the next section, we describe the
gravitational source and the Einstein-Maxwell field equations.
Section \textbf{3} is devoted to matching conditions. We formulate
the dynamical equations in section \textbf{4} and the derivation of
the transport equation and their coupling with the dynamical
equations are presented in section \textbf{5}. The last section
contains the conclusion of the paper.

\section{Interior Matter Distribution and the Field Equations}

We take non-static cylindrically symmetric as an interior metric in
the co-moving coordinates in the form
\begin{eqnarray}\label{1}
&&ds^{2}_{-}=-A^{2}dt^{2}+B^{2}dr^{2}+C^{2}(d{\theta}^{2}+dz^{2}),\\
&&-\infty<t<\infty,\quad 0\leqslant {r} <\infty,\quad
0\leqslant\theta\leqslant{2}\pi,\quad -\infty<z<+\infty\nonumber
\end{eqnarray}
where $A$, $B$ and $C$ are functions of $t$ and $r$. Matter under
consideration is anisotropic fluid which undergoes dissipation in
the form of heat flux. The energy-momentum tensor for such a fluid
dissipating only at diffusion approximation, i.e., $\epsilon=0$ is
defined as (Herrera 2006))
\begin{equation}\label{2}
T_{\alpha\beta}=(\mu+P_{\bot})V_{\alpha}V_{\beta}+P_{\bot}g_{\alpha\beta}
+(P_{r}-P_{\bot})\chi_{\alpha}
\chi_{\beta}+V_{\alpha}q_{\beta}+V_{\beta}q_{\alpha},
\end{equation}
where $\mu,~P_{r},~P_{\bot},~q_{\alpha},~V_{\alpha}$ and
$\chi_{\alpha}$ are the energy density, the radial pressure, the
tangential pressure, heat flux, the four-velocity of the fluid and
the unit four-vector along the radial direction respectively. For
the metric (1), the four-vector velocity, heat flux and unit
four-vector along the radial direction are given by
\begin{equation*}\label{3}
V^{\alpha}=A^{-1}\delta^{\alpha}_{0},\quad
\chi^{\alpha}=B^{-1}\delta^{\alpha}_{1}, \quad
q^{\alpha}=B^{-1}q{\delta}^{{\alpha}}_{1},
\end{equation*}
which satisfy
\begin{eqnarray*}\label{3}
V^{\alpha}V_{\alpha}=-1,\quad \chi^{\alpha}\chi_{\alpha}=1,\quad
\chi^{\alpha}V_{\alpha}=0, \quad q^{\alpha}V_{\alpha}=0.
\end{eqnarray*}

We can write the electromagnetic energy-momentum tensor in the form
\begin{equation}\label{3}
T^{(em)}_{\alpha\beta}=\frac{1}{4\pi}\left(F_{\alpha}^{\gamma}F_{\beta\gamma}
-\frac{1}{4}F^{\gamma\delta}F_{\gamma\delta}g_{\alpha\beta}\right).
\end{equation}
The Maxwell equations are given by
\begin{eqnarray}\label{4}
F_{\alpha\beta}&=&\phi_{\beta,\alpha}-\phi_{\alpha,\beta},\\\label{5}
{F^{\alpha\beta}}_{;\beta}&=&4\pi J^{\alpha},
\end{eqnarray}
where $F_{\alpha\beta}$ is the Maxwell field tensor, $\phi_{\alpha}$
is the four potential and $J_{\alpha}$ is the four current. Since
the charge is at rest with respect to the co-moving coordinate
system, thus the magnetic field is zero. Consequently, the four
potential and the four current will become
\begin{equation}\label{6}
\phi_{\alpha}=\phi{\delta^{0}_{\alpha}},\quad J^{\alpha}=\sigma
V^{\alpha},
\end{equation}
where $\phi=\phi(t,r)$ is an arbitrary function and
$\sigma=\sigma(t,r)$ is the charge density.

For the interior spacetime, using Eq.(\ref{6}), the Maxwell field
equations take the following form
\begin{eqnarray}\label{8}
\phi''-\left(\frac{A'}{A}+\frac{B'}{B}-2\frac{C'}{C}\right){\phi'}&=&{4
\pi }{\sigma}AB^{2},
\\\label{9}
{\dot{\phi}}'-\left(\frac{\dot{A}}{A}+\frac{\dot{B}}{B}-2\frac{\dot{C}}{C}\right){\phi'}&=&0,
\end{eqnarray}
where dot and prime represent the partial derivatives with respect
to $t$ and $r$ respectively. Integration of Eq.(\ref{8}) implies
that
\begin{equation}\label{10}
\phi'=\frac{2 sAB}{C^{2}},
\end{equation}
where $s\left(r\right)=2{\pi}{\int^{r}_{0}}\sigma BC^{2}dr$ is the
total charge distributed per unit length of the cylinder and is the
consequence of law of conservation of charge, $J^\mu_{; \mu}=0$.
Obviously Eq.(\ref{9}) is identically satisfied by Eq.(\ref{10}).

The Einstein field equations,
$G_{\alpha\beta}=8\pi(T_{\alpha\beta}+T^{(em)}_{\alpha\beta})$, for
the metric (1) can be written as
\begin{eqnarray}\label{12}
8{\pi}(T_{00}+T^{(em)}_{00})&=&8{\pi}{\mu}A^{2}+\frac{4s^{2}
{A^{2}}}{{C^{4}}}\nonumber\\
&=&\frac{\dot{C}}{C}\left(2\frac{\dot{B}}{B}+\frac{\dot{C}}{C}\right)
+\left(\frac{A}{B}\right)^{2}\left(-2\frac{C''}{C}
+2\frac{B'C'}{BC}-(\frac{C'}{C})^2\right),\nonumber\\\\
\label{13} 8{\pi}(T_{01}+T^{(em)}_{01})&=&8{\pi}q=\frac{2}{AB}
\left(\frac{\dot{C'}}{C}-\frac{\dot{B}C'}{BC}-\frac{\dot{C}A'}{CA}\right),\\
\label{14} 8{\pi}(T_{11}+T^{(em)}_{11})&=&8{\pi}P_{r}B^{2}
-\frac{4s^{2}B^{2}}{C^{4}}\nonumber\\
&=&-\left(\frac{B}{A}\right)^{2}\left(2\frac{\ddot{C}}{C}+\left(\frac{\dot{C}}{C}\right)^{2}
-2\frac{\dot{A}\dot{C}}{AC}\right)+\left(\frac{C'}{C}\right)^{2}+2\frac{A'C'}{AC},\nonumber\\
\end{eqnarray}
\begin{eqnarray}\label{15}
8{\pi}(T_{22}+E_{22})&=&8{\pi}P_{\bot}C^{2}
+\frac{4s^{2}}{C^{2}}\nonumber\\
&=&-\left(\frac{C}{A}\right)^{2}\left(\frac{\ddot{B}}
{B}+\frac{\ddot{C}}{C}-\frac{\dot{A}}{A}\left(\frac{\dot{B}}{B}+\frac{\dot{C}}{C}\right)
+\frac{\dot{B}\dot{C}}{BC}\right)\nonumber\\
&+&\left(\frac{C}{B}\right)^{2}\left(\frac{A''}
{A}+\frac{C''}{C}-\frac{A'}{A}\left(\frac{B'}{B}-\frac{C'}{C}\right)-\frac{B'C'}
{BC}\right).
\end{eqnarray}
The C-energy for the cylindrically symmetric spacetime is defined by
(Thorne 1965)
\begin{eqnarray}\label{16}
E=\frac{1}{8}(1-l^{-2}\nabla^{a}\tilde{r}\nabla_{a}\tilde{r}),
\end{eqnarray}
The circumference radius $\rho$, specific length $l$ and areal
radius $\tilde{r}$ can be defined as
\begin{eqnarray*}\label{2.1.16}
\rho^{2}=\xi_{(1)a}\xi^{a}_{(1)},\quad
l^{2}=\xi_{(2)a}\xi^{a}_{(2)},\quad \tilde{r}={\rho}l,
\end{eqnarray*}
where
$\xi_{(1)}=\frac{\partial}{\partial{\theta}},~\xi_{(2)}=\frac{\partial}{\partial{z}}$
are Killing vectors and $E$ represents the gravitational energy per
unit specific length of the cylinder.

The specific energy of the cylinder (Poisson 2004) analogous to
Misner and Sharp energy for the spherical symmetry in the interior
region with the contribution of electromagnetic field can be written
as follows
\begin{eqnarray}\label{17}
E'=\frac{l}{8}+\frac{C}{2}\left(\frac{\dot{C}^{2}}{A^{2}}
-\frac{C'^{2}}{B^{2}}\right)+\frac{s^{2}}{2C}.
\end{eqnarray}
We would like to mention here that this energy is also analogous to
Taub's mass function in the plane symmetric spacetime (Sharif and
Rehmat 2010 ).

\section{Junction Conditions}

In this section, we assume that the $3D$ timelike boundary surface
${\Sigma}$ splits the two $4D$ cylindrically symmetric spacetimes
$V^+$ and $V^-$. The metric which describes the internal region
$V^-$ is given by Eq.(\ref{1}) while for the representation of
exterior region $V^+$, a metric in the retarded time coordinate is
considered. If $M(\nu)$ and $Q(\nu)$ are mass and charge,
respectively, in retarded time then the corresponding cylindrically
symmetric spacetime given by (Chao-Guang 1995) will take the form
\begin{equation}\label{1j}
ds^{2}_{+}=-(\frac{-2M(\nu)}{R}+\frac{Q^{2}(\nu)}{R^{2}})d\nu^{2}-2dRd\nu+R^{2}(d\theta^2+dz^2),
\end{equation}
where $\nu$ is the retarded time coordinate. We can write the
induced metric for the hypersurface $\Sigma$ in the following form
\begin{equation}\label{2j}
(ds^{2})_{\Sigma}=-{d{\tau}}^2+y^{2}(\tau)(d\theta^{2}+dz^{2}),
\end{equation}
where $\xi^{i}\equiv(\tau,\phi,z)~(i=0,2,3)$ represent the intrinsic
coordinates of $\Sigma$.

The Darmois junction conditions (Darmois, 1927) can be stated as
follows:
\begin{itemize}
\item The continuity of the first fundamental form over the
hypersurface $\Sigma$ i.e.,
\begin{equation}\label{3j}
(ds^{2})_{\Sigma}=(ds^{2}_{-})_{\Sigma}=(ds^{2}_{+})_{\Sigma}.
\end{equation}
\item  The continuity of the second fundamental form over the
hypersurface $\Sigma$
\begin{equation}\label{4j}
[K_{ij}]=K^{+}_{ij}-K^{-}_{ij}=0.
\end{equation}
\end{itemize}
Here, $K^{\pm}_{ij}$ is the extrinsic curvature given by
\begin{equation}\label{5j}
K^{\pm}_{ij}=-n^{\pm}_{\sigma}(\frac{{\partial}^2\chi^{\sigma}_{\pm}}
{{\partial}{\xi}^i{\partial}{\xi}^j}+{\Gamma}^{\sigma}_{{\mu}{\nu}}
\frac{{{\partial}\chi^{\mu}_{\pm}}{{\partial}\chi^{\nu}_{\pm}}}
{{\partial}{\xi}^i{\partial}{\xi}^j}),\quad({\sigma},
{\mu},{\nu}=0,1,2,3).
\end{equation}
where $n^{\pm}_{\sigma}$ are the components of outward unit normal
to the hypersurface in the coordinates $\chi^{{\pm}\mu}$.

We can write the equations of hypersurface as follows
\begin{eqnarray}\label{6j}
h_{-}(t,r)&=&r-r_{\Sigma}=0,\\
h_{+}(\nu,R)&=&R-R_{\Sigma}(\nu)=0,\label{7j}
\end{eqnarray}
where $r_{\Sigma}$ is a constant. Using above equations, we have the
interior and exterior spacetimes on $\Sigma$ as follows
\begin{eqnarray}\label{8j}
(ds^{2}_{-})_{\Sigma}&=&-A^{2}(t,r_{\Sigma})dt^{2}+C^{2}(t,r_{\Sigma})(d\theta^{2}+dR_{\Sigma}(\nu){2}).\\
\label{9j}
(ds^{2}_{+})_{\Sigma}&=&-[(\frac{-2M(\nu)}{R_{\Sigma}(\nu)}+\frac{Q^2(\nu)}{R_{\Sigma}(\nu)^2})
+\frac{2dR_{\Sigma}(\nu)}{d\nu}]d\nu^{2}\nonumber\\
&+&R^{2}_{\Sigma}(d\theta^2+dz^2).
\end{eqnarray}
The continuity of the first fundamental form gives
\begin{eqnarray}\label{10j}
R_{\Sigma}(\nu)&=&C(t,r_{\Sigma}),\\ \label{11j}
\frac{dt}{d\tau}&=&\frac{1}{A},\\ \label{12j}
\frac{d\nu}{d\tau}&=&[(\frac{-2M(\nu)}{R_{\Sigma}}+\frac{Q^2(\nu)}{R_{\Sigma}^2})
+\frac{2dR_{\Sigma}}{d\nu}]^{\frac{-1}{2}}.
\end{eqnarray}
Now we consider the second fundamental form over $\Sigma$. For this
purpose, we need the outward unit normals to $\Sigma$ using
Eqs.(\ref{6j}) and (\ref{7j})
\begin{eqnarray}\label{13j}
n^{-}_{a}&=&B(0,1,0,0),
\\\label{14j}
n^{+}_{a}&=&
\left(\frac{-2M(\nu)}{R_{\Sigma}}+\frac{Q^2(\nu)}{R_{\Sigma}^2}
+\frac{2dR_{\Sigma}}{d\nu}\right)^{\frac{-1}{2}}(-\frac{d{R_{\Sigma}}}{d\nu},1,0,0).
\end{eqnarray}

The non-zero components of the extrinsic curvature $K^{\pm}_{ij}$
are
\begin{eqnarray}\label{15j}
K^{-}_{00}&=&-\left(\frac{A'}{AB}\right)_{\Sigma},\\ \label{41}
\label{16j}
K_{00}^{+}&=&[\frac{d^{2}\nu}{d\tau^{2}}(\frac{d\nu}{d\tau})^{-1}-
(\frac{M}{R^2}-\frac{Q^2}{R^3})(\frac{d\nu}{d\tau})]_{\Sigma}.\\
\label{17j}
K^{-}_{22}&=& K^{-}_{33}=\left(\frac{CC'}{B}\right)_{\Sigma},\\
\label{18j}
 K_{22}^{+}&=&[R\frac{dR}{d\tau}+(\frac{Q^2}{R}-2M)
\frac{d\nu}{d\tau}]_{\Sigma}=K_{33}^{+}.
\end{eqnarray}
The continuity of the extrinsic curvature components yields
\begin{eqnarray}\label{19j}
[\frac{d^{2}\nu}{d\tau^{2}}(\frac{d\nu}{d\tau})^{-1}-
(\frac{M}{R^2}-\frac{Q^2}{R^3})(\frac{d\nu}{d\tau})]_{\Sigma}
=-\left(\frac{A'}{AB}\right)_{\Sigma}\\\label{20j}
[R\frac{dR}{d\tau}+(\frac{Q^2}{R}-2M)
\frac{d\nu}{d\tau}]_{\Sigma}=\left(\frac{CC'}{B}\right)_{\Sigma}
\end{eqnarray}
Using Eqs.(\ref{10j})-(\ref{12j}), (\ref{12}) and (\ref{13}) in
Eqs.(\ref{19j}) and (\ref{20j}), it follows that
\begin{eqnarray}
{E'-M}\overset{\Sigma}{=}\frac{l}{8} \Leftrightarrow
s\overset{\Sigma}{=}Q,
\\\
q\overset{\Sigma}{=}P_{r}-\frac{3s^2}{2C^4}.
\end{eqnarray}
The first equation indicates that the difference between two masses
is equal to $\frac{l}{8}$ as shown in the adiabatic case (Sharif and
Fatima 2011). This is due to the least unsatisfactory definition of
Thorne C-energy (Thorne 1965). The second equation describes a
relation between heat flux, radial pressure and charge over the
hypersurface ${\Sigma}$. It is obvious from this equation that for
uncharged radiating fluid, radial pressure and heat flux are equal
over the boundary of the collapsing cylinder.

\section{The Dynamical Equations}

Here we derive the dynamical equations for non-adiabatic charged
anisotropic fluid. The energy-momentum conservation,
$(T^{\alpha\beta}+T^{{(em)}{\alpha\beta}})_{;\beta}=0$, implies that
\begin{eqnarray}\label{18}
\left(T^{\alpha\beta}+{T^{(em)}}^{\alpha\beta}\right)_{;\beta}V_{\alpha}=&-&\frac{\dot{\mu}}{A}
-\frac{\dot{B}}{AB}(\mu+P_{r})-\frac{2\dot{C}}{AC}(\mu+P_{\bot})\nonumber\\
&-&\frac{2q}{B}(\frac{A'}{A}+\frac{C'}{C})-\frac{q'}{B} =0
\end{eqnarray}
and
\begin{eqnarray}\label{19}
\left(T^{\alpha\beta}+{T^{(em)}}^{\alpha\beta}\right)_{;\beta}\chi_{a}
&=&\frac{1}{B}P_{r}'+\frac{A'}{AB}(\mu+P_{r})
+\frac{\dot{q}}{A}+\frac{2q}{A}(\frac{\dot{B}}{B}+\frac{\dot{C}}{C})\nonumber\\
&+&\frac{2C'}{BC}(P_{r}-P_{\bot}) -\frac{ss'}{\pi BC^{4}}=0
\end{eqnarray}
Following Misner and Sharp formalism (Misner and Sharp 1964), we
discuss the dynamics of the collapsing system. We introduce proper
time derivative as well as the proper radial derivative constructed
from the circumference radius of a cylinder inside $\Sigma$ as
(Sharif and Fatima 2011)
\begin{equation}\label{20}
D_{T}=\frac{1}{A}\frac{\partial}{\partial{t}},\quad
D_{R}=\frac{1}{R'}\frac{\partial}{\partial{r}},\quad R=C.
\end{equation}

The fluid velocity in the case of collapse can be defined as
\begin{equation}\label{23}
U=D_{T}(R)=D_{T}(C)<0.
\end{equation}
Consequently, we can write Eq.(\ref{17}) as
\begin{equation}\label{24}
\tilde{E}=\frac{C'}{B}=\left[U^{2}+\frac{s^{2}}{C^{2}}-\frac{2}{C}
\left(E'-\frac{1}{8}\right)\right]^{1/2}.
\end{equation}
Using Eqs.(\ref{13}), (\ref{14}), (\ref{17}) and (\ref{20}), the
time rate of change of the C-energy turns out to be
\begin{equation}\label{26}
D_{T}E'=-4{\pi}R^{2}\left[\left(P_{r}-\frac{1}{32{\pi}R^{2}}\right)U
+ q \tilde{E}\right]+\frac{3s^{2}U}{2R^{2}}.
\end{equation}
This equation represents the variation of total energy inside the
collapsing cylinder. Since $U<0$, the first term on the right hand
side of this equation will increase the energy of the system
provided that the factor within the round brackets is positive. The
second term in the square brackets due to negative sign describes
the outflow of energy in the form of radiation during the collapse.
For the collapsing cylinder containing the same species of the
charges, the third term will decrease the energy of the system as
$\frac{3s^2}{2R^2}$ plays the role of Coulomb repulsive force and
$U<0$.

Similarly, using Eqs.(\ref{12}), (\ref{13}), (\ref{17}) and
(\ref{20}), we obtain
\begin{equation}\label{27}
D_{R}E'=4{\pi}R^{2}({\mu}+q\frac{U}{\tilde{E}})+\frac{l}{8}
+\frac{s}{R}D_{R}s+\frac{3s^{2}}{2R^{2}}.
\end{equation}
This equation gives the variation of energy between the adjacent
cylindrical surfaces inside the matter distribution. The first term
is the energy density of the fluid element along with heat flux
contribution. Since $U<0$, the heat flux factor decreases the energy
of the system during the collapse of cylinder. The term
$\frac{l}{8}$ comes from the definition of C-energy and the
remaining terms are due to the electromagnetic field.

Using Eqs.(\ref{14}), (\ref{17}), (\ref{23}) and (\ref{24}), we can
obtain the acceleration $D_{T}U$ of a collapsing matter inside
$\Sigma$
\begin{equation}\label{28}
 D_{T}U=-\frac{1}{R^{2}}\left(E'-\frac{l}{8}\right)
-4\pi{R}P_{r}+\frac{\tilde{E}A'}{AB} +\frac{5s^{2}}{2R^{3}}.
\end{equation}
Inserting the value of $\frac{A'}{A}$ from Eq.(\ref{28}) into
Eq.(\ref{19}), it follows that
\begin{eqnarray}\label{29}
(\mu+P_{r})D_{T}U=&-&(\mu+P_{r})
\left[\frac{1}{R^{2}}(E'-\frac{l}{8})+4\pi P_{r}R
-\frac{5s^{2}}{2R^{3}}\right]\nonumber\\
&-&\tilde{E}\left[D_{T}q+\frac{4qU}{R}+2qG\right]\nonumber\\
&-&\tilde{E}^2\left[D_{R}P_{r} +2(P_{r}-P_{\bot}
)\frac{1}{R}-\frac{s}{{\pi}R^{4}}D_{R}s\right],
\end{eqnarray}
where $G=\frac{1}{A}(\frac{\dot{B}}{B}-\frac{\dot{C}}{C})\neq0$ for
simplicity. Now the complete dynamics of the system is described by
Eq.(\ref{29}). The system will evolve radially outward or inward
according to $D_{T}U<0$ or $D_{T}U>0$. Thus the terms in
Eq.(\ref{29}) contributing negatively, favors the collapse while the
other contribution prevents the collapse. If both of these cancel
each other then there will be a hydrostatic equilibrium. Since the
left hand side of Eq.(\ref{29}) represents force, so the factor
$\mu+P_{r}$ refers to an inertial mass density independent of charge
and heat flux contributions. The first and third terms on the right
hand side represents the gravitational force. The second term
represents the heat flux contribution which seems to leave the
system (due to negative sign) through the outward radially directed
streamlines. Being in the same direction of pressure, it supports
the pressure and would prevent the collapse.

The term $(\mu+P_{r})[\frac{1}{R^{2}}(E'-\frac{l}{8})+4\pi P_{r}R
-\frac{5s^{2}}{2R^{3}}]$ represents the gravitational force. The
factor within the first square brackets shows the effects of
specific length and the electric charge on the active gravitational
mass term $(\mu+P_{r})$. The third term has three main
contributions, i.e., the first is the pressure gradient which is
negative, the second is the local anisotropy of the fluid which will
be negative for $P_{r}<P_{\bot}$ and the third is the
electromagnetic field term. For an isotropic pressure, the second
contribution will be vanished. Further, following Di Prisco et al.
(2007), it can be found that the third term contributes negatively
for $\frac{s}{R}>D_{R}s $. Thus the third square brackets, under
these conditions with negative sign, contributes positively by
reducing an attractive behavior of force appearing on left hand side
of the equation. Since the attractive force is decreased, so the
third term prevents the gravitational collapse of the cylinder.

\section{The Transport Equation}

The transport equation predicts the processes of mass, heat and
momentum transfer during the dynamics of a realistic matter. The
transport equation for heat flux derived from the
M$\ddot{u}$ller-Israel-Stewart causal thermodynamic theory is given
by (Herrera 2006)
\begin{equation}\label{30}
\tau h^{ab}V^{c}q_{b;c}+q^{a}=-\kappa
h^{ab}(T_{,b}+a_{b}T)-\frac{1}{2}\kappa T^2(\frac{\tau V^{b}}{\kappa
T^2})_{;b}q^{a},
\end{equation}
where $h^{ab}=g^{ab}+V^{a}V^{b}$ is the projection tensor, $\kappa$
denotes thermal conductivity, $T$ is temperature, $\tau$ stands for
relaxation time which is the time taken by a perturbed system to
return into an equilibrium state and $a_{b}T$ is the Tolman inertial
term. Due to symmetry of the spacetime, the transport equation
reduces to the following form
\begin{equation}\label{31}
\tau \dot{q}=-\frac{1}{2}\kappa qT^2(\frac{\tau}{\kappa
T^2})^{\cdot}-\frac{1}{2}\tau
q(\frac{\dot{B}}{B}+2\frac{\dot{C}}{C}) -\frac{\kappa}{B}(TA)'-qA.
\end{equation}
Using Eqs.(\ref{20}) and (\ref{23}) in this equation, it follows
that
\begin{eqnarray}\label{32}
D_{T}q=&-&\frac{\kappa T^2q}{2\tau}D_{T}(\frac{\tau }{\kappa
T^2})-q[\frac{3U}{2 R}+{G}+\frac{1}{\tau}]-\frac{\kappa
\tilde{E}}{\tau }D_{R}T-\frac{\kappa
T}{\tau\tilde{ E}}\nonumber\\
&\times&D_{T}U-\frac{\kappa T}{\tau \tilde{E}}[E'+\frac{l}{8}+4\pi
P_{r}R^3-\frac{5s^{2}}{2R}]\frac{1}{R^2}.
\end{eqnarray}

In order to understand the effects of heat flux or dissipation on
collapsing process, we couple Eq.(\ref{32}) with dynamical
Eq.(\ref{29}). Thus the replacement of Eq.(\ref{32}) in
Eq.(\ref{29}), yields
\begin{eqnarray}\label{33}
(\mu+P_{r})(1-\alpha)D_{T}U&=&
(1-\alpha)F_{grav}+F_{hyd}+\frac{\kappa \tilde{E}^2}{\tau}
D_{R}T\nonumber\\&+&\tilde{E}[\frac{\kappa
T^2q}{2\tau}D_{T}(\frac{\tau}{\kappa
T^2})]-\tilde{E}q(\frac{5U}{2{R}} +G -\frac{1}{\tau}) ,
\end{eqnarray}
where $F_{grav},~F_{hyd}$ and $\alpha$ are given by the following
equations
\begin{eqnarray}\label{34}
F_{grav}&=&-(\mu+P_{r})[E'-\frac{C}{8}+4\pi
P_{r}{R}^3-\frac{5s^2}{2{R}}]\frac{1}{{R}^2},\\\label{35}
F_{hyd}&=&-\tilde{E}^2[D_{R}(P_{r})+\frac{2}{R}(P_{r}-P_{\perp})
-\frac{\mu_{0}^{2}sD_{R}s}{4\pi R^4}],
\\\label{36}
\alpha&=&\frac{\kappa T}{\tau}(\mu+P_{r})^{-1}.
\end{eqnarray}
For the physical interpretation of Eq.(\ref{33}), it can be observed
that the left hand side of this equation being the product of
inertial mass density $(\mu+P_{r}) (1-\alpha)$ and acceleration,
$D_{T}U$ represents the Newtonian force. Thus we can write
$F=(\mu+P_{r})(1-\alpha)D_{T}U$. It is clear that when
$\alpha\rightarrow1$, then $F\rightarrow0$, which means that there
is no inertial force and matter would experience the gravitational
attraction which causes the collapse. For $0<\alpha<1$, the inertial
mass density goes on decreasing while $1<\alpha$ indicates the
increase of inertial mass density. Of course, by the equivalence
principle, there would occur decrease and increase in the
gravitational mass. In this way, one can explicitly distinguish the
expanding and collapsing mechanism during the dynamics of
dissipative system.

Also, Eq.(\ref{33}) implies that gravitational force is affected by
the same factor but hydrodynamical force is independent of this
factor. Further, as long as $(\mu+P_{r})(1-\alpha)D_{T}U<0$, the
total Newtonian force of the system remains directed downward which
is the indication for the gravitational collapse and the converse is
true for the expansion. If there is a continuous change in $\alpha$
from a value greater than one to less than one and vice versa, then
there is a transition phase in the system and bouncing would occur.
This phenomenon causes the loss of energy from the system and hence
the collapsing cylinder with non-adiabatic source leads to the
emission of the gravitational radiations. On the basis of this fact,
the exterior of the collapsing cylinder is radiation zone which is
completely described by a spacetime in radiation coordinates like in
the present case.

\section{Outlook}

This paper deals with the effects of the charge and heat conduction
on the dynamics of cylindrical anisotropic fluid collapse. We have
extended the recent work of Sharif and Fatima (2011) to
non-adiabatic case for the transportation process of heat flux
during the dynamics of realistic matter collapse. For this purpose,
the non-viscous heat conducting anisotropic fluid with cylindrical
symmetry has been taken as the source of gravitation in the presence
of electromagnetic field. Using the Misner and Sharp formalism, the
dynamical equations are derived. We have found that during the
collapse of non-viscous heat conducting charged anisotropic fluid,
the radial heat flux and electric charge causes to reduce the energy
of the system.

For $q>0,~E'=\frac{C'}{B}>0$, Eq.(\ref{27}) yields that the second
term in the first square brackets is negative which indicates that
heat is emitting from the system. Thus the external region of such a
collapsing system, being non-vacuum, is defined by a cylindrical
geometry in the radiation (single null) coordinate. This prediction
is analogous to the Vaidya (1951) for spherically symmetric case.
The time evolution of the system, given by dynamical equation
(\ref{29}), indicates that the inertial mass density is independent
of heat flux and electric charge. The left hand side of this
equation corresponds to the Newtonian force of the system which is
decreased by the heat flux.

Since the collapse of a star is an irreversible process. The
transport process of such non-equilibrium objects and connection
between their dynamics and thermodynamics are important for the
better understanding of the problem. Thus using the
M$\ddot{u}$ller-Israel-Stewart causal thermodynamic theory, the
transport equation for the dissipative fluid has been formulated and
coupled to the dynamical equation. The coupled equation helps to
determine the influence of the heat flux over the dynamics of
collapsing cylinder. It has been found that in the coupled dynamical
Eq.(\ref{33}), the inertial as well as gravitational masses are
influenced by the factor $(1-\alpha)$. The role of $\alpha$ during
the dynamics of system can be explained as follows: For $\alpha$
tends to one, we get zero mass density. For $0<\alpha<1$, the
inertial and the gravitational mass density are decreased while for
$\alpha>1$, the gravitational force term becomes negative. This is
the case for the reversal of collapse. The conditions on $\alpha$
have been investigated for the bouncing behavior of the system.

\vspace{0.25cm}

{\bf Acknowledgment}

\vspace{0.25cm}

We would like to thank the Higher Education Commission, Islamabad,
Pakistan for its financial support through the {\it Indigenous Ph.D.
5000 Fellowship Program Batch-IV}.

\end{document}